\documentclass[%
 aip,
 amsmath,
 reprint,
 groupedaddress
]{revtex4-1}

\usepackage{graphicx}
\usepackage{dcolumn}
\usepackage{bm}

\usepackage[utf8]{inputenc}
\usepackage[T1]{fontenc}
\usepackage{amsmath,bookmark}
\usepackage{etoolbox}
\usepackage[dvipsnames]{xcolor}

\usepackage{stmaryrd}
\newcommand{\llrrbracket}[1]{\llbracket #1 \rrbracket}

\usepackage{newtxtext}
\usepackage{newtxmath}
\usepackage[scaled=0.82]{helvet}

\DeclareMathOperator*{\argmax}{arg\,max}

\hypersetup{
  colorlinks=true,
  urlcolor=RoyalBlue,
  linkcolor=RoyalBlue,
  citecolor=RoyalBlue
}

\makeatletter
\def\@email#1#2{%
 \endgroup
 \patchcmd{\titleblock@produce}
  {\frontmatter@RRAPformat}
  {\frontmatter@RRAPformat{\produce@RRAP{*#1\href{mailto:#2}{#2}}}\frontmatter@RRAPformat}
  {}{}
}%
\makeatother
\begin{document}

\title{\protect \LARGE Morphodynamics of lipid vesicles under osmotic forcing\vspace{0.3cm}}
\author{\protect\large Nicholas Broussinos}
\affiliation{ 
Department of Mechanical and Aerospace Engineering, University of California San Diego, \\ 9500 Gilman Drive, La Jolla, CA 92093, USA
}%
\author{David Saintillan$^*$\vspace{0.12cm}}%
 \email{dstn@ucsd.edu}
\affiliation{ 
Department of Mechanical and Aerospace Engineering, University of California San Diego, \\ 9500 Gilman Drive, La Jolla, CA 92093, USA
}%

\begin{abstract}\vspace{-0.1cm}
We investigate the dynamics of spherical lipid vesicles subjected to an osmotic shock using continuum simulations of a permeable, area-incompressible fluid membrane with bending elasticity and surface viscosity. The dynamics are governed by a single dimensionless parameter $\varGamma$, which measures the relative strength of osmotic pressure and bending forces. Osmotic forcing drives a three-stage evolution comprising rapid buckling and collapse, smoothing of the resulting wrinkles and ridges, and slow folding into an invaginated morphology resembling the precursor to experimentally observed vesicle-in-vesicle structures. Linear stability analysis predicts the wavelength and growth rate of the initial buckling instability and shows that the dimensionless buckling time scales as $\varGamma^{-2}$, in quantitative agreement with simulations. At longer times, the energy budget reveals a reversal in elastic energy transfer: bending energy accumulated during buckling is released during relaxation and dissipated by membrane viscosity. A scaling analysis of the folding dynamics predicts that the dimensionless folding time scales as $\varGamma^{-2/3}$, again in excellent agreement with simulations. These results provide a continuum description of the mechanisms and widely separated timescales governing osmotically driven membrane collapse and invagination.
\end{abstract}

\maketitle

\section*{\protect\Large Introduction}

\noindent Lipid bilayers are ubiquitous in biology, forming the membranes that surround cells and many intracellular organelles. These membranes exhibit a remarkable diversity of morphologies, ranging from nearly spherical vesicles to discocyte and stomatocyte shapes in red blood cells and the highly folded structures found in the inner mitochondrial membrane. Much of our understanding of these forms comes from curvature-elastic theories, which describe membrane shape as the result of minimizing an elastic energy subject to geometric constraints such as fixed area and volume \cite{helfrich_elastic_1973,seifert_shape_1991}. Such approaches successfully account for many experimentally observed morphologies, including the taxonomy of red blood cell shapes through mechanisms such as the bilayer-couple model \cite{lim_h_w_stomatocytediscocyteechinocyte_2002}. However, while equilibrium membrane shapes are relatively well understood, the dynamical pathways by which membranes undergo large-scale shape transformations remain far less explored.

One striking example is the response of a vesicle to an osmotic shock. Because lipid bilayers are permeable to water, a difference in solute concentration across the membrane can drive substantial volume changes. Under sufficiently strong hypertonic forcing, a spherical vesicle loses stability, expels fluid, and develops wrinkles and folds that initiate a complex sequence of morphological transitions leading to highly invaginated shapes
\cite{vanhille-campos_modelling_2021,zong_fissionable_2017,TL2021}. 
Experiments and molecular dynamics simulations have shown that osmotic buckling may ultimately produce vesicle-in-vesicle structures through a sequence of collapse, elastic relaxation, and topology-changing fission events \cite{bernard_raspberry_2002,zong_fissionable_2017,zong_deformation_2018}. 
These transformations have been observed across a broad range of vesicle sizes, from roughly 100\,nm to 10\,$\mu$m \cite{zong_fissionable_2017,vanhille-campos_modelling_2021}. Despite these observations, the continuum-scale mechanisms governing the instability, the emergence of highly collapsed intermediate states, and the long-time evolution toward invaginated morphologies remain poorly understood.

At continuum scales, lipid bilayers behave as viscous fluid surfaces endowed with bending elasticity. Their dynamics therefore couple in-plane lipid flows to out-of-plane shape deformations, leading to a moving two-dimensional Stokes problem on an evolving manifold.
Lipid bilayers occupy an intermediate regime between liquid drops, which lack bending elasticity, and solid shells, which resist in-plane deformation. Governing equations have been formulated for bending-elastic, surface-viscous interfaces, drawing from the study of both fluid films and solid shells \cite{hu_continuum_2007,arroyo_relaxation_2009,sahu_irreversible_2022}, but simulating viscous flow on a deforming surface remains challenging. Large deformations during osmotic collapse generate highly curved and self-contacting configurations that are difficult to capture using traditional membrane simulation methods. To address these challenges, Zhu \textit{et al.}\cite{zhu_mem3dg_2022,zhu_stokes_2025} used the machinery of discrete differential geometry to develop an efficient solver based on Onsager’s variational principle, which extends Rayleigh’s principle of least energy dissipation to nonlinear irreversible systems and is powerful for the description of soft matter \cite{onsager_reciprocal_1931,doi_onsagers_2011}. Minimization of the resulting dissipation functional, known as the Rayleighian, yields the membrane velocity and tension fields at each instant of time.  

\begin{figure*}[!t]
\includegraphics[width=0.95\textwidth]{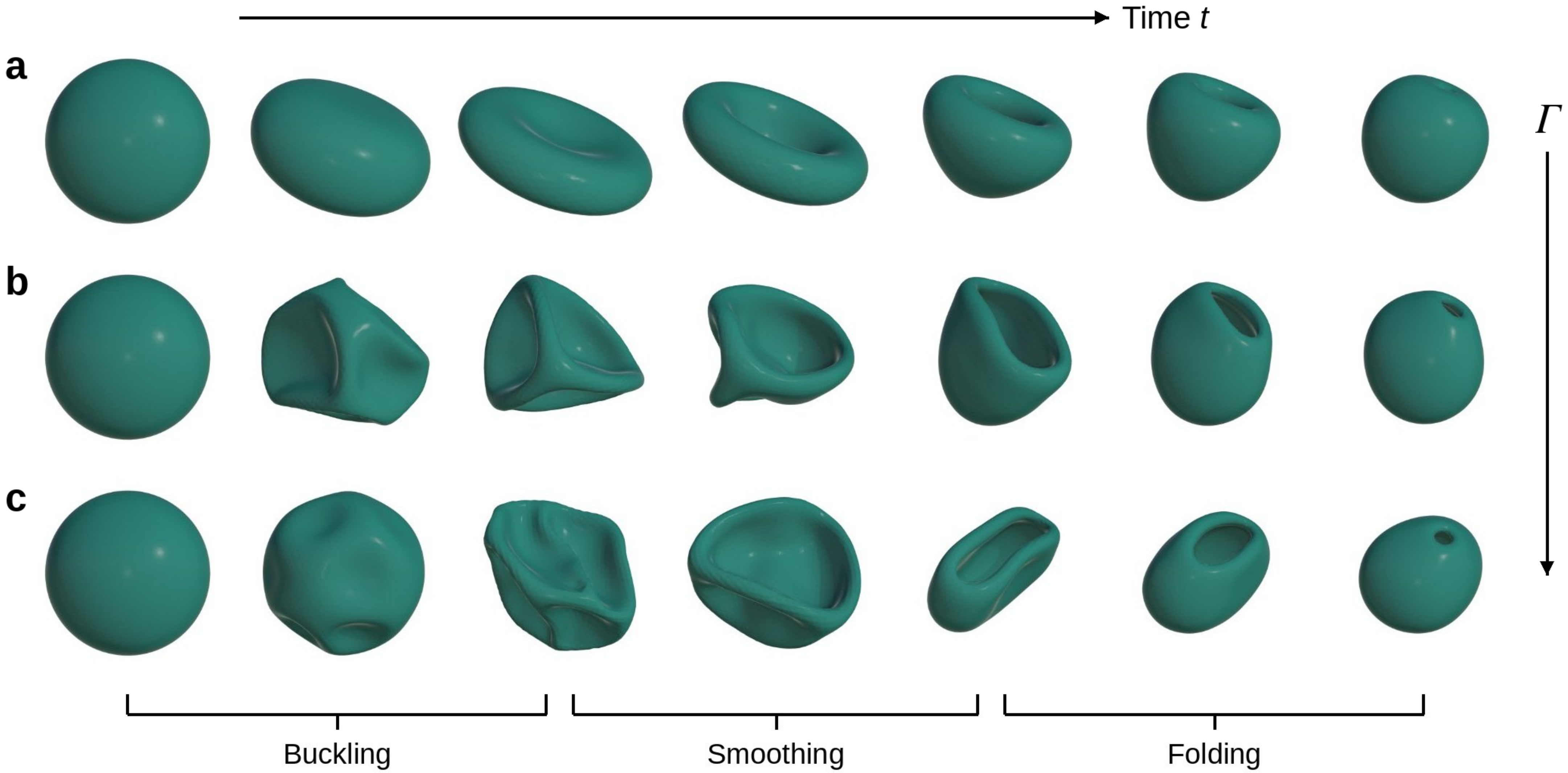}\vspace{-0.0cm}
\caption{\textbf{Snapshots from three simulations with increasing values of $\varGamma$}. Time increases from left to right. Snapshots are taken at the initial time, the moment of contact, the moment when smooth relaxation begins, and final simulation time, along with intermediate frames. Frames are not equally spaced in time, and are not synchronized between rows; see Supplemental Videos 1 to 3 for movies showing the dynamics. \textbf{(a)}~At $\varGamma=0.2$, the vesicle buckles into an axisymmetric discocyte shape. \textbf{(b)} At $\varGamma=2.1$, it buckles into an axially periodic shape with 4 pockets and 6 ridges. \textbf{(c)} At $\varGamma=3$, it buckles into a disordered, wrinkled shape. Note the increasing eccentricity of the roughly elliptical opening during smoothing, and the convergence toward a similar deeply invaginated, pre-fission morphology in all three cases despite vastly different intermediate dynamics. }\vspace{-0.1cm}  
\label{fig:snapshots}
\end{figure*}

In this work, we perform continuum simulations of osmotic buckling by building on the formulation of Zhu \textit{et al.} \cite{zhu_mem3dg_2022,zhu_stokes_2025}. We consider the idealized limit in which the osmotic pressure difference remains uniform and constant throughout the dynamics, and assume negligible hydraulic resistance to solvent permeation through the membrane.  We find that osmotic buckling drives vesicles through a sequence of well-separated dynamical regimes: rapid wrinkling, slow relaxation into highly collapsed states, and eventual formation of morphologies resembling vesicle-in-vesicle configurations. Since topological changes are not included in the present continuum framework, our simulations terminate in an invaginated precursor state that resembles the experimentally observed pre-fission morphology. Scaling arguments are put forth to explain the various timescales involved in this process based on the results of linear stability analysis and on a three-term dominant balance suggested by the energetic dynamics. Our results provide a mechanistic framework for understanding osmotically driven membrane remodeling in a broad range of biological and synthetic membrane systems.

\section*{\protect\Large Results}

We perform numerical simulations of the dynamics of an initially spherical vesicle of radius $R$ subjected to a constant and uniform osmotic pressure jump $\llbracket p \rrbracket$. The membrane is modeled as a closed manifold $\mathcal{M}$ with in-plane viscous and out-of-plane elastic behavior (surface viscosity $\mu$, bending modulus $\kappa_b$). It is surface incompressible, a constraint that is satisfied by the spatially dependent surface tension $\lambda$, which can be interpreted as a Lagrange multiplier. The governing equations and numerical methods extend the work of Zhu \textit{et al.} \cite{zhu_mem3dg_2022,zhu_stokes_2025} and are presented in the Methods section. We discuss results in dimensionless form, where lengths are scaled by $R$ and time by the elastoviscous timescale $t_c=\mu R^2/\kappa_b$. Under this scaling, the dynamics of the system are entirely governed by a single dimensionless osmotic forcing parameter,
\begin{equation}
\varGamma = \frac{\llrrbracket{p}R^3}{16\pi\kappa_b},
\end{equation}
which compares the magnitudes of the osmotic pressure jump and of bending forces in the initially spherical vesicle. 

Our aim is to characterize the dynamics of buckling and subsequent relaxation as a function of $\varGamma$. Snapshots from three simulations with increasing values of $\varGamma$ are shown in Fig.~\ref{fig:snapshots}; also see Supplemental Videos 1 to 3 for corresponding movies showing the dynamics. Across the range of forcing strengths explored here, the dynamics can be divided into three successive stages. On short timescales, buckling under the applied pressure jump results in the collapse of the initially spherical vesicle, giving rise to morphologies whose characteristic length scales decrease as the pressure is increased. At small $\varGamma$, the vesicle develops an axisymmetric biconcave shape reminiscent of a red blood cell. As $\varGamma$ is increased, the buckled vesicle takes on a more wrinkled appearance with increasingly smaller and numerous features, including dimples, ridges and folds. On intermediate timescales, following the initial shape collapse, the vesicle enters a smoothing regime, in which the sharp ridges that formed during buckling get smoothed out and dissipate, yielding an invaginated cup-shaped morphology. On the longest timescales, the membrane undergoes a spontaneous invagination that produces two nearly concentric membrane compartments connected by a narrow neck. The neck radius decreases continuously throughout the simulations, suggesting progression toward the experimentally observed pre-fission state. We begin by examining the initial buckling instability and then analyze the intermediate relaxation and long-time invagination dynamics separately.  

\begin{figure*}[t!]
\includegraphics[width=0.92\textwidth]{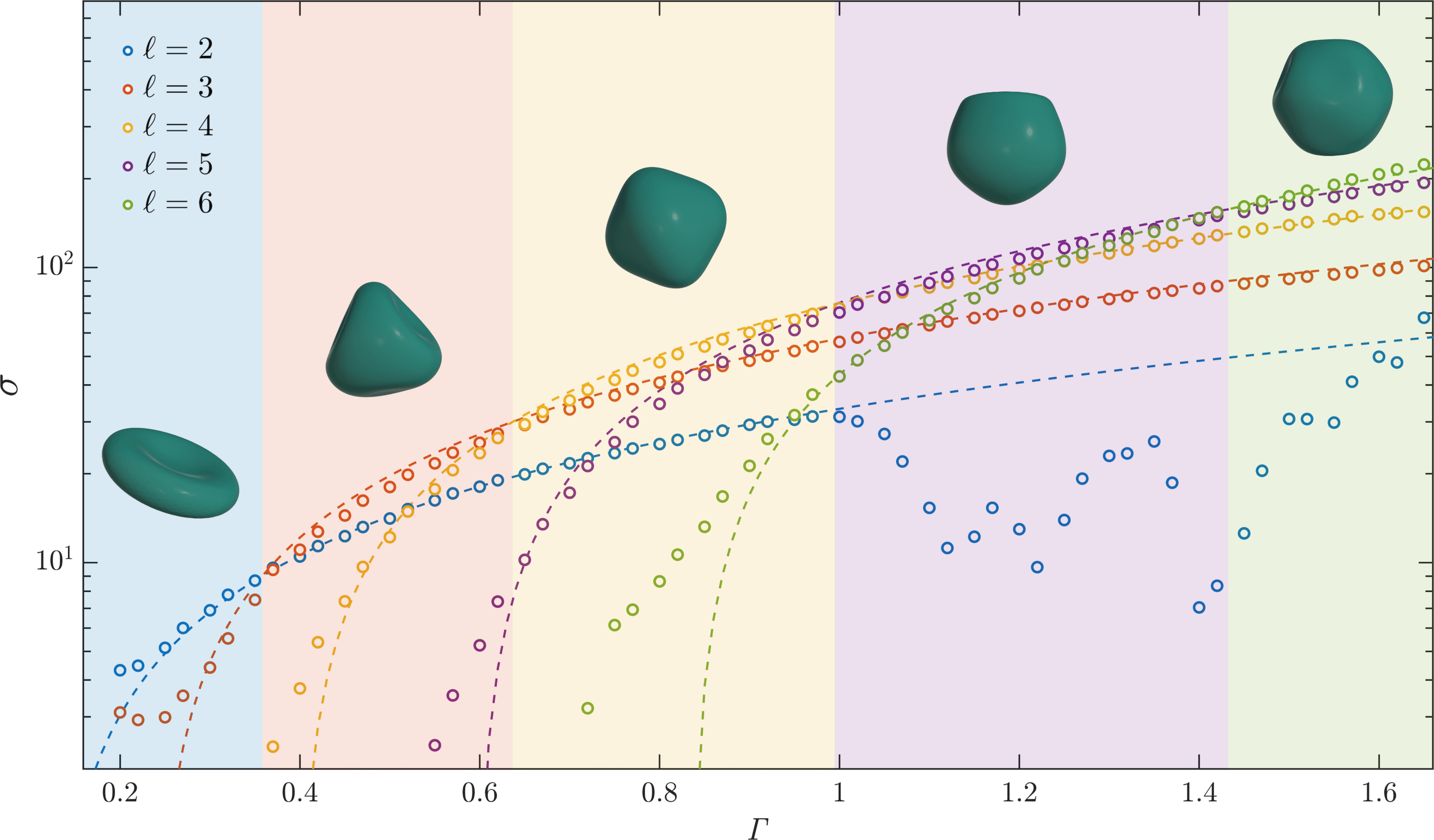}\vspace{-0.0cm}
\caption{\textbf{Linear stability predictions and comparison to numerical results.} The theoretical growth rate of Eq.~(\ref{eq:sigma}) is plotted versus $\varGamma$ for various values of the degree $\ell$ (dashed lines), and compared to numerical growth rates extracted from nonlinear simulations (symbols, see Methods for details). The colored background highlights the most unstable mode at a given $\varGamma$ value. Typical shapes from nonlinear simulations shortly after buckling are shown to illustrate the dominant mode of instability in each region.} \label{fig:linear}\vspace{-0.0cm}
\end{figure*}

\subsection*{\protect\large Short times: pressure-induced elastic buckling}

The initial collapse of the vesicle is governed by a linear buckling instability of the spherical shape (see Methods for details). A linear stability analysis due to Sahu\cite{sahu_irreversible_2022} predicts that perturbations proportional to spherical harmonics $Y_{\ell m}$ with degree $\ell$ and order $m$ grow at a dimensionless rate
\begin{equation}
\sigma = \frac{\ell(\ell + 1)}{4}\left(8\pi\varGamma - \frac{\ell(\ell + 1)}{2}\right).\label{eq:sigma}
\end{equation}
The osmotic pressure jump destabilizes the spherical shape, while bending penalizes curvature and therefore suppresses sufficiently short-wavelength perturbations. Because the undeformed vesicle is perfectly spherical, all spherical harmonics of degree $\ell$ have identical growth rates, and become unstable when $\varGamma \ge \ell(\ell+1)/16\pi$. Upon increasing $\varGamma$, more modes become unstable, and at each $\varGamma$ one mode is predicted to grow most rapidly.  The dominant unstable mode also increases sequentially with $\varGamma$, starting with the quadrupolar mode $\ell=2$. The competition between pressure and bending therefore selects a characteristic buckling wavelength that decreases as $\varGamma$ increases.

We test these linear stability predictions against the early-time dynamics in nonlinear simulations in Fig.~\ref{fig:linear}. Simulations were performed for varying $\varGamma$, starting from a spherical shape perturbed by small-amplitude white noise so as to excite all unstable spherical harmonic modes. For quantitative comparison, the evolving membrane shapes were projected onto spherical harmonics using a least-squares regression, and numerical growth rates for each mode were computed from the time evolution of the power spectrum (see Methods). Excellent agreement is observed for the dominant unstable modes, while the weaker modes exhibit modest deviations from linear theory. Consistent with the linear theory, increasing $\varGamma$ shifts the dominant instability toward larger values of $\ell$, explaining the progressively finer wrinkling observed during the initial collapse.
This agreement confirms that the early-time dynamics are governed by linear buckling and establishes the initial conditions from which the strongly nonlinear collapse develops.
We next examine the nonlinear evolution responsible for wrinkle formation, relaxation, and invagination.

\begin{figure*}[t!]
\includegraphics[width=\textwidth]{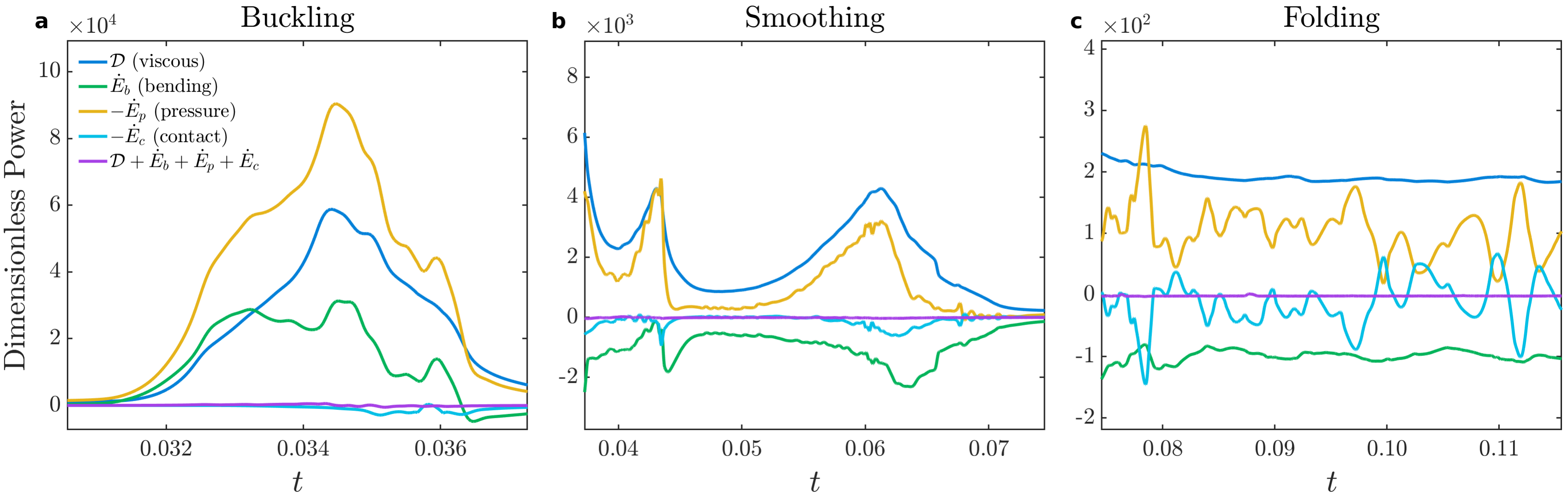}
\caption{\textbf{Power budget during the buckling, smoothing and folding regimes in a simulation with $\varGamma=1$.} Contributions to the instantaneous power balance during the \textbf{(a)} buckling, \textbf{(b)} smoothing, and \textbf{(c)} folding regimes. Shown are the viscous dissipation rate $\mathcal{D}$, rate of change of bending energy $\smash{\dot{E}_b}$, power supplied by the applied osmotic pressure jump $\smash{-\dot{E}_p}$, and power supplied by contact forces $\smash{-\dot{E}_c}$. With this sign convention, the energy balance reads $\smash{\mathcal{D}+\dot{E}_b=-\dot{E}_p-\dot{E}_c}$. During buckling, pressure work is partitioned between bending-energy storage and viscous dissipation. During smoothing and folding, $\smash{\dot{E}_b<0}$, indicating release of stored bending energy, which together with pressure work is dissipated by membrane viscosity. The energy-balance residual $\smash{\mathcal{D}+\dot{E}_b+\dot{E}_p+\dot{E}_c}$ (purple) remains close to zero throughout the dynamics.}  
\label{fig:power}
\end{figure*} 

\subsection*{\protect\large Nonlinear dynamics and energy transfer}

Following the initial linear buckling, the membrane undergoes a sequence of deformations that can be divided into three distinct regimes: nonlinear buckling resulting in vesicle collapse and wrinkle formation, pressurized smoothing of the wrinkles, followed by slow folding and invagination. To identify mechanisms controlling these stages, we examine the instantaneous power budget for a representative simulation at $\varGamma=1$ in Fig.~\ref{fig:power}. From the variational formulation, the mechanical energy balance reads
\begin{equation}
    \dot{E}_b+\dot{E}_p+\dot{E}_c+{\mathcal{D}}=0,
\end{equation}
where $\mathcal{D}=2\mu \int_\mathcal{M} \mathbf{E}\boldsymbol{:}\mathbf{E}\,\mathrm{d}A$ is the rate of viscous dissipation in the membrane. This relation provides a direct measure of how work performed by the osmotic pressure is partitioned between elastic deformation, contact interactions, and viscous dissipation.

During nonlinear buckling [Fig.~\ref{fig:power}(a)], the osmotic pressure provides the dominant source of mechanical power. This input is divided between the creation of increasingly curved membrane structures, reflected in the growth of bending energy, and viscous dissipation associated with the rapid surface flow. Remarkably, these two contributions remain comparable over much of the collapse, indicating that neither bending elasticity nor surface viscosity can be treated as a perturbative correction during buckling.

Self-contact marks the end of the rapid collapse and a qualitative change in the energy budget. Whereas bending energy increases during buckling, $\dot{E}_b$ becomes negative during the subsequent relaxation. There are two qualitatively different intervals of the relaxation dynamics, henceforth referred to as smoothing and folding. During smoothing [Fig.~\ref{fig:power}(b)], the highly curved ridges and wrinkles generated during buckling are smoothed out, releasing stored bending energy. Together with continued work by the pressure load, this released elastic energy is dissipated by membrane viscosity. The smoothing dynamics are energetically characterized by sharp spikes in the constituent work rates, which coincide with the rapid relaxation of localized, highly curved features by a combination of pressure and bending forces. 

At the end of the smoothing stage, the membrane approaches an axisymmetric cup shape. At longer times [Fig.~\ref{fig:power}(c)], the dynamics enters a much slower folding regime in which the membrane reduces its shape energy by folding into a configuration with two roughly concentric compartments connected by a narrowing neck. Bending energy continues to decrease slowly, while pressure and contact forces fluctuate as opposing membrane surfaces remain in close proximity. 
The neck does not reach a steady radius but rather becomes infinitesimally small, at which point the simulations become unphysical as they are unable to resolve topological changes. 

Across all three regimes, a striking conclusion is that the power budget does not reduce to a simple balance between pressure and bending. During buckling, the work of pressure forces is divided between elastic energy storage and viscous dissipation, with the latter two contributions of comparable magnitude. After collapse, the direction of elastic energy transfer reverses: pressure work and the release of stored bending energy jointly feed viscous dissipation. Thus pressure forcing, bending elasticity, and surface viscosity remain coupled at leading order throughout the dynamics, despite the distinct mechanisms involved during buckling, smoothing, and folding.

\subsection*{\protect\large Deformation timescales}

A separation of timescales is observed between the various dynamical regimes; as the osmotic forcing is varied, the ratios between buckling time, smoothing time, and folding time vary as well. We denote the durations of buckling, smoothing, and folding by $t_b$, $t_s$ and $t_f$, respectively, and seek scaling relations for their dependence on $\varGamma$. Recall that dimensional analysis identifies the elastoviscous time $t_c=\mu R^2/\kappa_b$
as a characteristic dimensional timescale for the problem. 

The timescale for the buckling regime, defined as the duration of membrane collapse prior to self-contact, can be estimated from the growth rate of the dominant unstable mode. The linear growth rate in Eq.~(\ref{eq:sigma}) depends on both $\ell$ and $\varGamma$. For a fixed value of $\varGamma$, we define the dominant wavenumber as 
\begin{equation}
\ell_{d}(\varGamma) = \argmax_{\ell}\sigma (\ell,\varGamma).
\end{equation}
Treating $x=\ell(\ell+1)$ as a continuous variable, the dispersion relation is maximized at $x_d=\ell_d(\ell_d+1)=8\pi\varGamma$. For sufficiently large $\ell_d$, this gives $\ell_d\approx \sqrt{8\pi\varGamma}$, and, upon substitution into  Eq.~(\ref{eq:sigma}), a maximum growth rate $\sigma_{\mathrm{max}}\sim \varGamma^2$. The characteristic buckling time can then be estimated as
\begin{equation}
\frac{t_b}{t_c} \sim \sigma_{\mathrm{max}}^{-1}\sim \varGamma^{-2}.  \label{eq:bucklingtime}
\end{equation}

The smoothing regime is less amenable to a simple scaling argument because its geometry and dominant mechanisms change with $\varGamma$. At low-to-moderate forcing, smoothing primarily involves bending-driven relaxation of approximately periodic folds of wavelength $\lambda_s\sim R\varGamma^{-1/2}$, whereas at stronger forcing the collapsed surface develops localized ridges and valleys that relax through translation and coalescence. Moreover, pressure and bending can either cooperate or oppose one another in different regions of the membrane. We therefore do not expect smoothing to be governed by a single characteristic length scale or dominant balance.

During folding, the shape of the vesicle is much simpler and can be divided into two domains: one, corresponding to the two concentric nearly spherical caps, where contact forces balance the effect of osmotic pressure, and another, corresponding to the circular neck, where pressure forces are resisted primarily by bending. The neck region is approximately toroidal, with a minor radius $r_{\!f}$ set by the balance of pressure and bending forces,
\begin{equation}
\llrrbracket{p} \sim \kappa_br_{\!f}^{-3}. 
\end{equation}
This provides a simple scaling for the minor radius of the toroidal neck,
\begin{equation}
    \frac{r_{\!f}}{R}\sim \varGamma^{-1/3}.
\end{equation}
During folding, material is transported from the outer surface of the cup toward the invaginated region, while material already within the cup undergoes comparatively little displacement. The characteristic length scale of the surface flow is therefore $r_{\!f}$, giving a characteristic velocity $U\sim r_{\!f}/t_f$ and strain rate $\smash{E\sim U/r_{\!f}\sim t_{\!f}^{-1}}$. Since pressure and bending forces balance locally in the neck, both have characteristic magnitude $\smash{\kappa_b r_{\!f}^{-3}}$. The characteristic mechanical power density associated with neck motion therefore scales as $\smash{\kappa_b r_{\!f}^{-3}U}$. Balancing this with the viscous dissipation density gives 
\begin{equation}
    \frac{\kappa_{b\vphantom{f}}}{r_{\!f}^3}\frac{r_{\!f}}{t_{\!f}^{\vphantom{3}}} \sim \frac{\mu}{t_{\!f}^2},
\end{equation}
or
\begin{equation}
    t_f\sim \frac{\mu r_{\!f}^2}{\kappa_b}.
\end{equation}
Thus, the folding dynamics occur on the local elastoviscous timescale associated with the neck radius rather than the vesicle radius. Using $t_c=\mu R^2/\kappa_b$ together with $r_{\!f}/R\sim\varGamma^{-1/3}$ then yields
\begin{equation}
\frac{t_f}{t_c}\sim \left(\frac{r_{\!f}}{R}\right)^2 \sim \varGamma^{-2/3}.  \label{eq:foldingtime}
\end{equation}

\begin{figure}[]
\includegraphics[width=0.98\columnwidth]{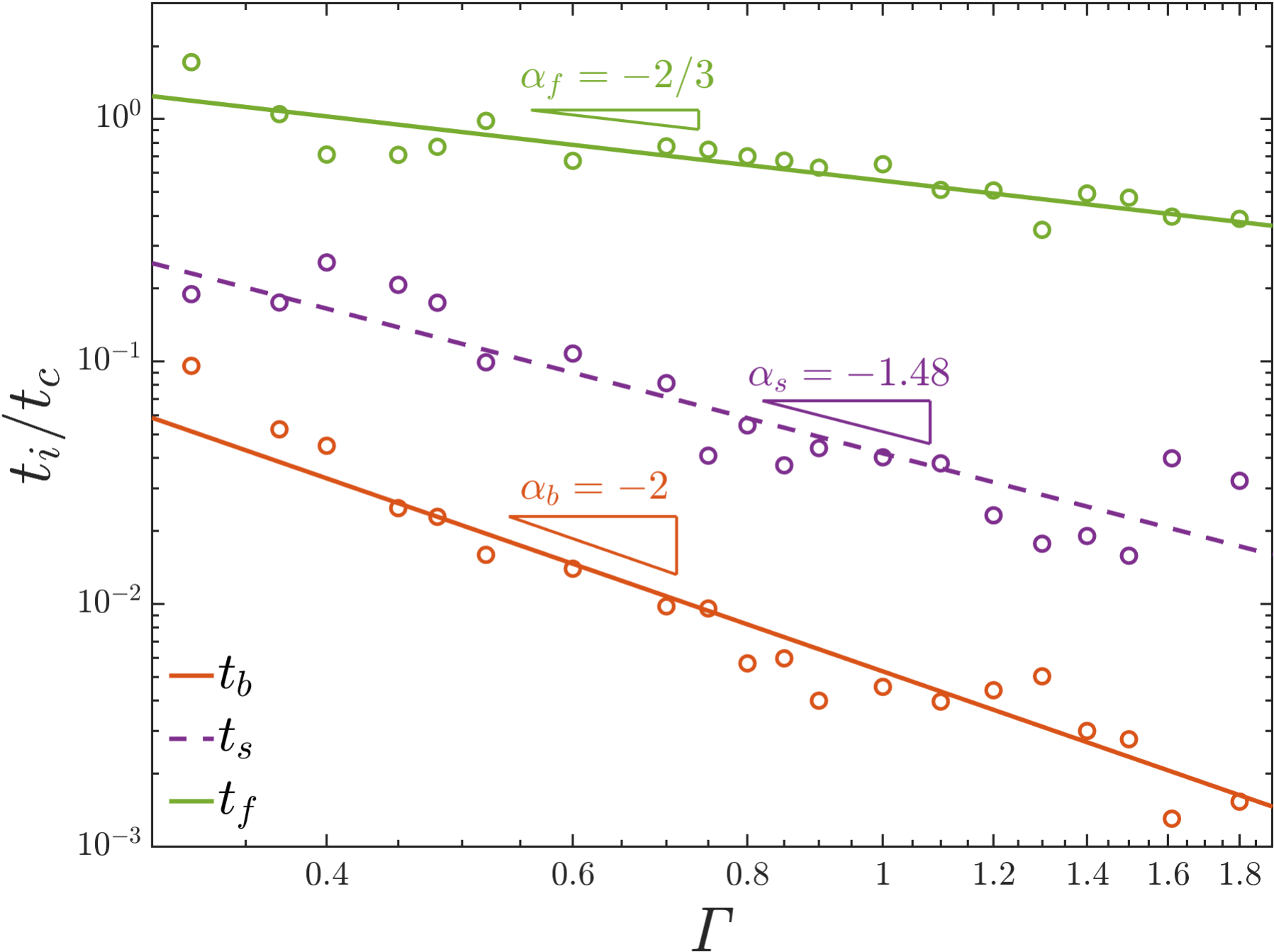}
\caption{\textbf{Scaling of deformation timescales with osmotic forcing.} Dimensionless buckling ($t_b/t_c$), smoothing ($t_s/t_c$), and folding ($t_f/t_c$) times as functions of $\varGamma$, extracted from numerical simulations. Solid lines show the theoretical predictions $t_b/t_c\sim\varGamma^{-2}$ and $t_f/t_c\sim\varGamma^{-2/3}$. A power-law fit to the smoothing time, shown by the dashed line, yields an exponent of $-1.48$, suggesting an empirical scaling close to $t_s/t_c\sim\varGamma^{-3/2}$.
} \label{fig:timescales}
\end{figure}

Numerically, we extract the buckling, smoothing and folding times from the energetic dynamics of each simulation, like those shown in Fig.~\ref{fig:power}. When plotted versus $\varGamma$ in Fig.~\ref{fig:timescales}, the timescales display a power-law dependence that can be compared to our analytical predictions. The buckling and folding exponents are in excellent agreement with the theoretical predictions of $-2$ and $-2/3$, respectively. A numerical fit of the smoothing time yields a power-law exponent of $-1.48$, suggesting an empirical scaling close to $\varGamma^{-3/2}$.

\section*{\protect\Large Discussion}

We have analyzed the dynamic shape transformations of lipid vesicles subjected to a uniform pressure load as an idealized model of osmotic shock. Vesicles are modeled as area-incompressible fluid membranes with bending elasticity and surface viscosity, in the limits of high permeability and negligible bulk hydrodynamic dissipation. Under these assumptions, the dynamics are governed by a single dimensionless osmotic forcing parameter $\varGamma$, measuring the relative strength of pressure and bending forces. Our simulations reveal that osmotic collapse of a fluid vesicle is organized into three well-separated dynamical regimes: a pressure-driven buckling instability, relaxation of the highly curved structures generated during collapse, and slow folding toward an invaginated state. These three regimes involve distinct mechanics and timescales. The initial buckling results from a linear instability that selects the wavelength of deformation and occurs on a dimensionless timescale of order $\varGamma^{-2}$. The smoothing regime is strongly nonlinear with no clear dominant balance and an empirically observed timescale scaling approximately as $\varGamma^{-3/2}$. Finally, the slow folding regime, occurring on a timescale of order $\varGamma^{-2/3}$, is governed by the localized mechanics at the neck, whose geometry depends on the balance of pressure and bending forces. A separation between rapid collapse and much slower relaxation was also observed in previous molecular dynamics simulations \cite{vanhille-campos_modelling_2021}; here, the continuum description identifies the distinct mechanical origins and scaling behavior of these timescales.

One striking finding of our analysis, revealed by the instantaneous power budget, is the prominent role of surface viscosity. During vesicle collapse, the work of pressure forces contributes to both elastic energy storage and viscous dissipation. During the smoothing and folding regimes that follow collapse, stored bending energy is released and, together with continued pressure work, dissipated by surface viscosity. Thus, the post-buckling evolution cannot be understood from membrane elasticity alone, and surface viscosity remains a leading-order contribution throughout the dynamics. This feature distinguishes fluid lipid membranes from elastic shells \cite{H2016} and capsules \cite{KK2014}, which store energy through in-plane elastic deformation. While the initial pressure-driven instability shares qualitative features with elastic-shell buckling, the subsequent dynamics are fundamentally different: lipid bilayers possess no static shear elasticity and instead accommodate large tangential rearrangements through viscous surface flow. Surface viscosity therefore does not merely set the rate of an otherwise elastic relaxation, but participates at leading order in the nonlinear force and energy balances.

Over the range of $\varGamma$ investigated, we found that the vesicle at long times evolves towards a roughly axisymmetric invaginated morphology consisting of two nearly concentric surfaces connected by an increasingly narrow neck. Similar shapes have been reported in a variety of experiments\cite{zong_fissionable_2017,zong_deformation_2018,TL2021} as well as molecular dynamics simulations\cite{vanhille-campos_modelling_2021}, where continued evolution can lead to membrane fission and the formation of a vesicle-in-vesicle configuration. The continuum dynamics therefore provide a mechanical route to the pre-fission morphology observed experimentally, while the final topology-changing event lies outside the scope of the present model. Our simulations terminate when the neck radius approaches the mesh resolution. In physical vesicles, continued neck thinning will eventually invalidate the smooth-surface continuum description, as molecular-scale effects associated with finite bilayer thickness and lipid rearrangements become important, ultimately enabling topological transitions such as membrane fission.

The computational model presented here relies on a number of approximations and simplifications, which may affect the dynamics in some regimes. First, we modeled the osmotic shock in terms of a spatially and temporally constant pressure jump across the membrane. This approximation corresponds to the limit of a large, well-mixed external reservoir whose osmotic contribution dominates that of the internal solution throughout collapse. More generally, for a fixed amount of impermeable internal solute, the internal solute concentration increases as the vesicle deflates, thereby reducing the osmotic pressure difference. This could potentially arrest collapse or alter the long-time dynamics. Second, we have considered the limit of high permeability, where hydraulic resistance to solvent permeation is negligible. At finite permeability, hydraulic dissipation introduces an additional timescale associated with solvent transport and could modify both collapse and relaxation. Finally, we neglected the effect of bulk hydrodynamics, an approximation that is justified when the Saffman--Delbr\"uck length $\ell_{\mathrm{SD}}$ is large compared to the vesicle radius. Including inner and outer bulk fluids would introduce $R/\ell_{\mathrm{SD}}$ as an additional dimensionless parameter and could modify the deformation timescales. The present model thus isolates the membrane-dominated, highly permeable limit in which the dynamics are controlled by a single forcing parameter $\varGamma$, and it establishes a basis for incorporating finite permeability, bulk hydrodynamics, and topological changes in future work. 

More broadly, our results show how osmotic forcing, bending elasticity, and surface viscosity organize membrane collapse across widely separated time and length scales. The rapid initial dynamics are controlled by linear mode selection on the scale of the vesicle, whereas the slow late-time evolution is governed by nonlinear mechanics localized near a narrowing neck. The resulting scaling laws connect these distinct regimes within a single continuum framework and demonstrate how different physical mechanisms govern the rapid onset of buckling and the much slower progression toward strongly collapsed and invaginated states.

\section*{\protect\Large Methods}

\subsection*{\protect\large Governing equations for membrane dynamics}

We model the lipid vesicle as a closed two-dimensional manifold $\mathcal{M}$ embedded in three-dimensional space, following the formulation of Zhu \textit{et al.}\cite{zhu_stokes_2025} Let $\mathbf{n}$ denote the outward unit normal to the membrane surface and $\mathbf{S}=\nabla\mathbf{n}$ its second fundamental form, where $\nabla$ denotes the covariant derivative on $\mathcal{M}$. The mean and Gaussian curvatures are defined as $H=\tfrac{1}{2}\mathrm{tr}\,\mathbf{S}$ and $K=\det\,\mathbf{S}$, respectively. The velocity of a material point $\mathbf{r}\in\mathcal{M}$ is decomposed into tangential and normal components according to
\begin{equation}
\dot{\mathbf{r}}=\mathbf{u}+u_n\mathbf{n},
\end{equation}
where $\mathbf{u}$ is the in-plane velocity and $u_n$ is the normal velocity. The corresponding surface rate-of-strain tensor is
\begin{equation}
\mathbf{E}
=\frac{1}{2}(\nabla\mathbf{u}+\nabla\mathbf{u}^{\mathsf T})
+u_n\mathbf{S}.
\end{equation}

The membrane is modeled as an incompressible Newtonian surface fluid with surface viscosity $\mu$. Its surface stress tensor is given by
\begin{equation}
\mathbf{\Sigma}=2\mu\mathbf{E}-\lambda\mathbf{I},
\end{equation}
where the spatially varying membrane tension $\lambda$ enforces local area incompressibility,
\begin{equation}
\mathrm{tr}\,\mathbf{E}
=\nabla\boldsymbol{\cdot}\mathbf{u}+2Hu_n=0.
\label{eq:incompressibility}
\end{equation}
In addition to its in-plane viscous response, the membrane resists bending according to the Helfrich energy
\begin{equation}
E_b=2\kappa_b\int_{\mathcal{M}}H^2\,\mathrm{d}A,
\label{eq:bending_energy}
\end{equation}
where $\kappa_b$ is the bending modulus. The vesicle is driven by a spatially uniform and temporally constant pressure jump $\llbracket p\rrbracket$ across the membrane, with associated potential energy
\begin{equation}
E_p=\llbracket p\rrbracket V,
\label{eq:pressure_energy}
\end{equation}
where $V$ is the instantaneous enclosed volume. We additionally introduce a contact energy $E_c$ to prevent membrane self-intersection during highly collapsed configurations. The explicit form and numerical implementation of $E_c$ are described further below.

Following Zhu \textit{et al.},\cite{zhu_stokes_2025} the membrane dynamics are obtained from Onsager's variational principle. The viscous dissipation rate associated with membrane deformation is
\begin{equation}
\mathcal{D}(\mathbf{u},u_n)
=2\mu\int_{\mathcal{M}}\mathbf{E}\boldsymbol{:}\mathbf{E}\,\mathrm{d}A,
\end{equation}
and the corresponding dissipation potential is $\Phi=\mathcal{D}/2$. 
The incompressibility constraint is imposed through the Lagrange multiplier $\lambda$, leading to the Rayleighian
\begin{equation}
\mathcal{R}(\mathbf{u},u_n,\lambda)
=
\Phi(\mathbf{u},u_n)
+\frac{\mathrm{d}}{\mathrm{d}t}\left(E_b+E_p+E_c\right)
-\int_{\mathcal{M}}\lambda\,\mathrm{tr}\,\mathbf{E}\,\mathrm{d}A.
\label{eq:Rayleighian}
\end{equation}
At each instant, the membrane velocity and tension are obtained by requiring stationarity of $\mathcal{R}$ with respect to $\mathbf{u}$, $u_n$, and $\lambda$. Variation with respect to $\lambda$ enforces the local area-incompressibility constraint Eq.~(\ref{eq:incompressibility}), while variations with respect to the membrane velocity yield the local force balance,
\begin{align}
\nabla\boldsymbol{\cdot}\mathbf{\Sigma}
-(\mathbf{S}\boldsymbol{:}\mathbf{\Sigma})\mathbf{n}=
\mathbf{f}_b+\mathbf{f}_p+\mathbf{f}_c.
\label{eq:force_balance}
\end{align}
Together, the two terms on the left-hand side represent the surface divergence of the membrane stress in three-dimensional space, with the first giving its tangential component and the second the normal contribution arising from surface curvature. The bending, pressure, and contact forces on the right-hand side follow from variations of the corresponding energy functionals. In particular,
\begin{align}
\mathbf{f}_b
&=-\frac{\delta E_b}{\delta\mathbf{r}}
=2\kappa_b\left[\Delta H+2H(H^2-K)\right]\mathbf{n},\\
\mathbf{f}_p
&=-\frac{\delta E_p}{\delta\mathbf{r}}
=\llbracket p\rrbracket\mathbf{n},
\label{eq:pressure_force}
\end{align}
where $\Delta$ denotes the Laplace--Beltrami operator on $\mathcal{M}$; the contact force $\mathbf{f}_c$ is specified below.

Equations~(\ref{eq:incompressibility}) and (\ref{eq:force_balance}) describe a two-dimensional Stokes flow on an evolving curved surface, coupled to normal deformation through bending, pressure, and contact forces. In the present model, hydrodynamic stresses exerted by the surrounding bulk fluids and dissipation associated with transmembrane water transport are neglected. The physical limits associated with these approximations, together with the interpretation of $\llbracket p\rrbracket$ as an osmotic pressure difference, are discussed in the following subsection. A detailed derivation of the surface Stokes equations and definitions of the associated differential operators can be found in Zhu \textit{et al.}\cite{zhu_stokes_2025} 

\subsection*{\protect\large Osmotic forcing and dimensionless parameters}

The system of governing equations introduced above relies on several physical assumptions that we rationalize here. We first justify why osmotic forcing can be incorporated as an effective pressure jump $\llbracket p\rrbracket$ across the membrane in Eq.~(\ref{eq:pressure_energy}). For a semipermeable membrane, the volume flux through the membrane is described by the Kedem--Katchalsky relation,\cite{KK1958}
\begin{equation}
\mathbf{J}_v=-\beta\,\llbracket p-\Pi\rrbracket\,\mathbf{n},
\label{eq:flux}
\end{equation}
where $\beta$ is the hydraulic permeability, $p$ is the hydrostatic pressure in the bulk, and $\Pi=nk_BT$ is the osmotic pressure of a dilute solution, with $n$ the number density of osmotically active solute particles. We define the jump of a quantity $f$ as $\llbracket f\rrbracket=f_{\mathrm{out}}-f_{\mathrm{in}}$, so that Eq.~(\ref{eq:flux}) predicts solvent flow from high to low hydrostatic pressure and from low to high osmotic pressure. 
In the high-permeability limit, the difference $\llbracket p-\Pi\rrbracket$ required to sustain a finite permeation velocity is $O(\beta^{-1})$, while the leading-order hydrostatic pressure jump remains $\llbracket p\rrbracket\simeq\llbracket\Pi\rrbracket$.
Dissipation associated with solvent permeation is then negligible on the timescale of membrane deformation, and osmotic forcing can be represented as a mechanical pressure load through the potential $E_p=\llbracket p\rrbracket V$ in Eq.~(\ref{eq:pressure_energy}).

We further idealize this osmotic pressure difference as constant throughout the dynamics. The external solution is treated as a large, well-mixed reservoir, so that $n_{\mathrm{out}}$ remains constant during vesicle collapse. Conservation of impermeable solute inside the vesicle would generally cause $n_{\mathrm{in}}$ to increase as the enclosed volume decreases, thereby reducing the osmotic pressure difference. Here we consider the limit in which the external solute concentration dominates throughout the collapse, $n_{\mathrm{in}}(t)\ll n_{\mathrm{out}}$, such that $\llbracket\Pi\rrbracket\simeq n_{\mathrm{out}}k_BT$ remains approximately constant.

Another approximation of the model is the neglect of viscous stresses exerted by the surrounding bulk fluids. A characteristic membrane viscous force per unit area scales as $f_{\mathrm{mem}}\sim \mu U/R^2$ where $U$ is a characteristic velocity scale, whereas the viscous traction exerted by the bulk fluid with viscosity $\mu_b$ scales as $f_{\mathrm{bulk}}\sim \mu_b U/R$. Their ratio therefore scales as $f_{\mathrm{bulk}}/f_{\mathrm{mem}}\sim R/\ell_{\mathrm{SD}}$, where $\ell_{\mathrm{SD}}=\mu /\mu_b$ is the Saffman--Delbr\"uck length. In the limit $\ell_{\mathrm{SD}}\gg R$, membrane viscous stresses dominate over bulk viscous tractions, justifying their neglect in the present model. Experimentally reported values of $\ell_{\mathrm{SD}}$ for lipid membranes range from hundreds of nanometers to several micrometers,\cite{FDV2022} so this approximation is most appropriate for small vesicles and/or membranes with relatively large surface viscosity.

In the limit of large permeability and large Saffman-Delbr\"uck length considered here, the dynamics is simply governed by the competition of osmotic pressure and bending elasticity, with surface viscous stresses serving as a mechanism for energy dissipation. In this regime, the osmotic forcing parameter $\varGamma = \llrrbracket{p}R^3/(16\pi\kappa_b)$, comparing the magnitude of the osmotic pressure jump and bending forces in the undeformed spherical configuration, fully governs the dynamics. 

\subsection*{\protect\large Numerical implementation and self-contact}

The governing equations are solved numerically after discretization of the membrane surface into a triangular mesh. Rather than solving the local form Eq.~(\ref{eq:force_balance}) of the force balance, we employ a variational method that directly minimizes a discrete version of the Rayleighian of Eq.~(\ref{eq:Rayleighian}) while updating the membrane tension to satisfy area incompressibility. The reader is referred to Zhu \textit{et al.}\cite{zhu_stokes_2025} for a complete exposition of the discretization scheme and variational integrator. 

During membrane buckling, care must be taken to prevent unphysical self-intersection of the membrane surface. In our numerical simulations, this is achieved by means of the contact energy $E_c$ that enters the Rayleighian in Eq.~(\ref{eq:Rayleighian}). To enforce self-contact while taking into account the connectivity of the mesh, we employ the tangent-point energy of Sassen \textit{et al.}\cite{sassen_repulsive_2024} defined as 
\begin{align}
E_c &= \xi \int_{\mathcal{M}}\int_{\mathcal{M}} \frac{|\mathbf{n}(\mathbf{r})\boldsymbol{\cdot}(\mathbf{r}-\mathbf{r}')|^{\alpha}}{|\mathbf{r}-\mathbf{r}'|^{2\alpha}}\,\mathrm{d}A\,\mathrm{d}A',  \label{eq:tangentpoint}
\end{align}
where $\xi$ and $\alpha$ are user-defined parameters. This energy captures a power-law repulsion between pairs of points on the membrane with effective exponent $\alpha$. Unlike a purely distance-based repulsion, the tangent-point energy distinguishes neighboring points along a smooth surface from points that are close in Euclidean space but distant along the membrane. The normal projection in the numerator suppresses interactions between locally coplanar surface elements while strongly penalizing the approach of distinct portions of the membrane. The corresponding contact force is approximated from the variational derivative of the tangent-point energy,
\begin{align}
    &\mathbf{f}_c = - \frac{\delta E_c}{\delta \mathbf{r}} \label{eq:contactforce} \\
    &\,\,\approx-\xi \alpha \!\!\int_{\mathcal{M}}\!\!\!\!\!\frac{|\mathbf{n}(\mathbf{r})\cdot (\mathbf{r}-\mathbf{r}')|}{|\mathbf{r}-\mathbf{r}'|}\!\left[\frac{\mathbf{n}(\mathbf{r})}{\mathbf{n}(r)\cdot(\mathbf{r}-\mathbf{r}')}-\frac{2(\mathbf{r}-\mathbf{r}')}{|\mathbf{r}-\mathbf{r}'|^2}\right]\mathrm{d}A', \nonumber
\end{align}
where higher-order terms arising from the gradient of the surface normal are ignored with no impact on the repulsive character of the force.
Numerically, the integrals in Eqs.~(\ref{eq:tangentpoint}) and (\ref{eq:contactforce}) are performed via a vertex-based quadrature with barycentric area weights. 

The implementation of the tangent-point energy in our simulations presents two computational challenges. First, its numerical evaluation is costly due to the double integral over all pairs of points $(\mathbf{r},\mathbf{r}')$, resulting in an $O(N^2)$ computational cost with respect to mesh size. We remedy this by leveraging $k$-d tree data structures. Furthermore, because the tangent-point energy is used here solely to regularize short-range self-contact, interactions beyond a cutoff distance $r_c$ are neglected.

\begin{figure}[]
\includegraphics[width=\columnwidth]{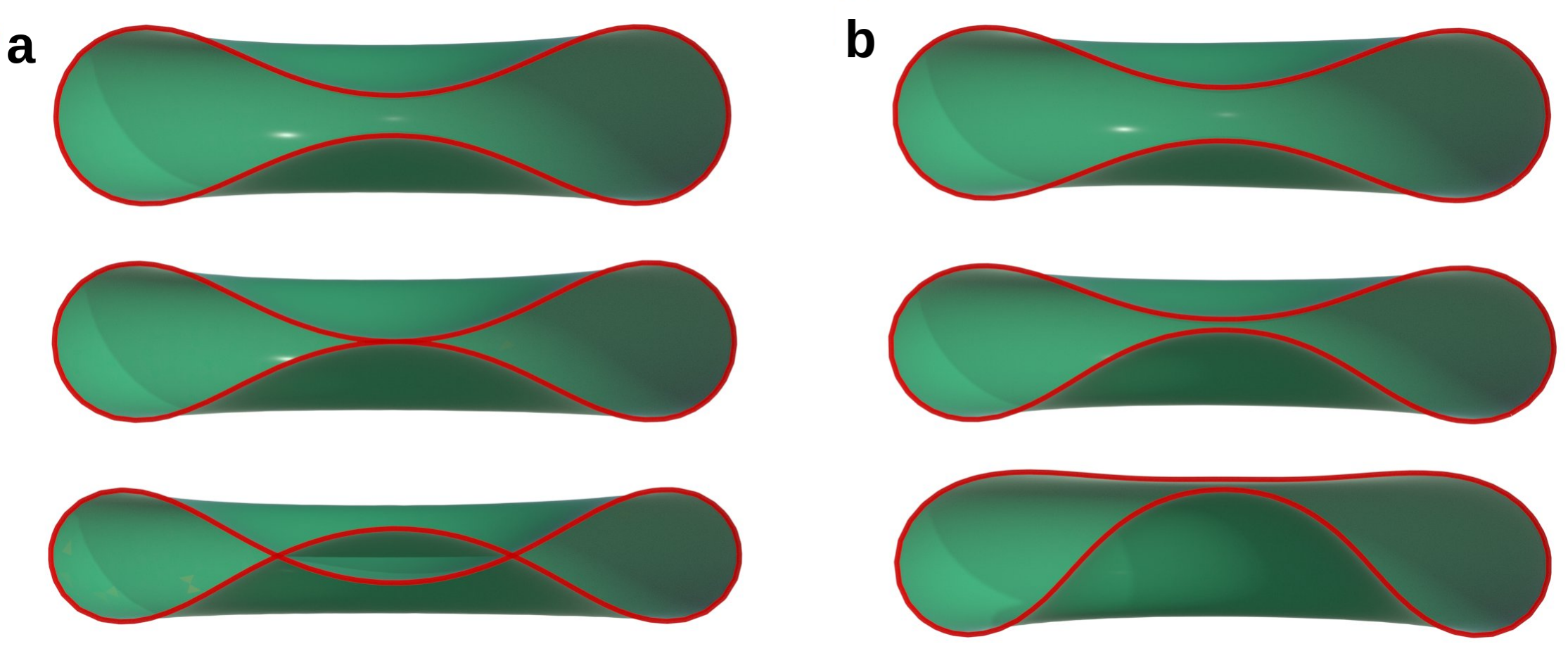}
\caption{\textbf{Effect of tangent-point repulsion on membrane self-contact.} Successive snapshots of a buckling vesicle \textbf{(a)} without and \textbf{(b)} with the tangent-point contact energy, with time progressing from top to bottom. Red curves show cross-sectional profiles through the membrane surface. Without contact interactions, opposing portions of the membrane meet and subsequently pass through one another, leading to unphysical self-intersection. With tangent-point repulsion, self-intersection is prevented; following first contact, the vesicle instead breaks its top--bottom reflection symmetry and continues to collapse while maintaining excluded volume. } 
\label{fig:contact}
\end{figure}

Second, due to the large separation of length scales between contact and the other membrane forces, the mesh fidelity required to properly resolve contact is extremely small. To alleviate this, we introduce an adaptive grid refinement algorithm for the efficient computation of the tangent-point gradient. We first perform a long-range ($\alpha = 3$) calculation of the gradient on the coarse grid to identify faces where repulsion may be relevant in the local force balance. Each flagged triangular face is temporarily subdivided into four subtriangles by introducing points at its edge midpoints, with positions and normals interpolated from the parent mesh. The gradient (with $\alpha = 4$) is then computed on the locally refined mesh and the vector field is retained at the vertices of the coarse mesh. This procedure increases the spatial resolution of the contact calculation without refining the underlying membrane mesh used to solve the mechanical problem. The result is an efficient detection algorithm for surface contact that yields an appropriate repulsive correction to the discrete momentum equation. Figure~\ref{fig:contact} contrasts the moment of self-contact with and without the tangent-point energy in the Rayleighian. The grid refinement technique can also be applied recursively for contexts where a particularly minute contact thickness must be resolved. 

All simulations used a time step of $\Delta t = 0.0012\,\varGamma^{-2}$. Long-time simulations were performed using a grid with 3230 points, while the analysis of the short-time growth rates used a finer grid with 12208 points. The magnitude of the tangent-point energy was set to $\xi=-8.89\times10^{-7}\varGamma$, with a cutoff radius of $r_c/R=0.15$ for the truncation of contact forces. 

\subsection*{\protect\large Linear stability analysis and spherical harmonic decomposition}

The linear stability of a spherical viscous membrane subject to a uniform pressure jump has been analyzed previously using scalar and vector spherical harmonics \cite{sahu_irreversible_2022}. Here we briefly summarize the results needed for comparison with our numerical simulations. Perturbations to the spherical base state are expanded as
\begin{align}
\begin{split}
\tilde{\mathbf{v}} =
\sum_{\ell = 0}^{\infty} \sum_{m = -\ell}^{+\ell}
\left[
v_{\ell m}^{s} \mathbf{Y}_{\ell m}^{s}
+ v_{\ell m}^{e} \mathbf{Y}_{\ell m}^{e}
+ v_{\ell m}^{m} \mathbf{Y}_{\ell m}^{m}
\right],\,\,\,\,\, \\
\tilde{r} = R+ \sum_{\ell=0}^{\infty} \sum_{m=-\ell}^{\ell} r_{\ell m} Y_{\ell m},\quad \tilde{\lambda} = \sum_{\ell=0}^{\infty} \sum_{m=-\ell}^{\ell} \lambda_{\ell m} Y_{\ell m},
\end{split}
\end{align}
where $Y_{\ell m}$ and $\mathbf{Y}_{\ell m}^{s,e,m}$ denote the scalar and vector spherical harmonics, respectively. Linearization of the governing equations and projection onto these modes yield decoupled evolution equations for the radial amplitudes,
\begin{equation}
\dot{r}_{\ell m} =\sigma_\ell r_{\ell m},
\end{equation}
with dimensionless growth rate
\begin{equation}
\sigma_\ell=
\frac{\ell(\ell+1)}{4}
\left[
8\pi\varGamma-\frac{\ell(\ell+1)}{2}
\right].
\label{eq:stability_methods}
\end{equation}
The growth rate is independent of \(m\), as expected from the rotational symmetry of the spherical base state. Equation~(\ref{eq:stability_methods}) implies that modes of degree \(\ell\) become unstable when \(\varGamma>\ell(\ell+1)/(16\pi)\). A complete derivation is given by Sahu \cite{sahu_irreversible_2022}.

To compare these predictions with our simulations, the instantaneous membrane shape is decomposed into scalar spherical harmonics by least-squares regression using the Spherical Harmonic Transform Library \cite{politis_microphone_2016}. We compute the degree-\(\ell\) power spectrum
\begin{equation}
S(\ell,t)=\frac{1}{2\ell+1}
\sum_{m=-\ell}^{\ell}|r_{\ell m}(t)|^2.
\end{equation}
In the linear regime, \(S(\ell,t)\propto\exp(2\sigma_\ell t)\); numerical growth rates are therefore obtained from linear fits of \(\log S(\ell,t)\) versus time, with \(\sigma_\ell\) equal to one half of the fitted slope.

\section*{\protect\Large Data availability}

The data are available from the authors upon request. 

\section*{\protect\Large Code availability}

The software is available from the authors upon request. 

\section*{\protect\Large References\vspace{-0.25cm}}

\section*{\protect\Large Acknowledgements}

We gratefully acknowledge support from the National Science Foundation (NSF) grants 2153520 and 2327243, and thank Cuncheng Zhu and Michael Shelley for fruitful discussions. 

\section*{\protect\Large Author contributions}

Both authors designed the research. N.B. developed the software, performed simulations, analyzed data, prepared figures, and wrote the manuscript. Both authors interpreted results and data. D.S. edited the manuscript.  

\section*{\protect\Large Competing interests}

The authors declare no competing interests.


\end{document}